# Two distinct superconducting regimes in $Ti_4Co_2O$ under pressures

Lifen Shi[1, 2, 3 =], Keyuan Ma[3, 4 =], Binbin Ruan[1, 2=], Zhen Wang[1,2], Pengtao Yang[1,2], Zhian Ren[1,2], Jianping Sun[1,2], Gang Li[1,2], Fabian O. von Rohr[4]*, Bosen Wang[1,2]* and Jinguang Cheng[1,2]*

[1]Beijing National Laboratory for Condensed Matter Physics and Institute of Physics, Chinese Academy of Sciences, Beijing 100190, China

[2]University of Chinese Academy of Sciences, Beijing 100190, China

[3]Max Planck Institute for Chemical Physics of Solids, 01187 Dresden, Germany

[4]Department of Quantum Matter Physics, University of Geneva, CH-1211 Geneva

= These authors contributed equally to this work.

*Corresponding authors: bswang@iphy.ac.cn; Fabian.VonRohr@unige.ch; jgcheng@iphy.ac.cn

**Abstract**

We report on the pressure dependence of superconducting transition temperature $T_c$ and upper critical field $B_{c2}(0)$ through electrical transport of the $\eta$-carbide $Ti_4Co_2O$ superconductor (eg., the superconducting transition temperature $T_c \approx 2.5$ K and the $B_{c2}(0) \approx 7.2$ T $\approx 2.9T_c$). We find that the $T_c$ exhibits non-monotonic pressure dependence: it rises monotonically at first with a pressure coefficient of $dT_c/dP \approx$ 0.034 K/GPa, but rapidly decreases around 10-20 GPa, and then increases with the $dT_c/dP \approx 0.023$ K/GPa, up to ≈ 4.31 K at 69.7 GPa. Concurrently, the $B_{c2}(0)$ exhibits a dome shaped pressure dependence, with its maximum at 5 GPa of almost twice the value at ambient pressure, exceeding the weak-coupling Pauli paramagnetic limit $B_p^{BCS}$ throughout the whole pressure range. By comparing the normal-state and superconducting properties, we identify two distinct superconducting regimes, with a low-pressure superconducting phase (< 12 GPa) characterized by an enhanced $B_{c2}(0)$ values and Fermi-liquid normal-state electrical transport (the exponent $n \approx 2$), and a high-pressure superconducting phase (> 20 GPa) with a monotonically increased $T_c$ and an enhancement in phonon scatterings (the exponent $n \approx 4$). Room-temperature synchrotron X-ray diffraction indicates that there is no structural transition up to 55.8 GPa, which gives a relatively large bulk modulus of ~ 192 GPa in comparison with other alloy superconductors. First-principles calculations suggest that the nonmonotonic $T_c$ maybe closely related to the evolution of the density of states of $Ti_4Co_2O$ upon compression, which is different from those of isostructural superconductors $Ti_4Ir_2O$ and $Nb_4Rh_2C_{1-\delta}$. Our results show that even in the $\eta$-carbide $Ti_4Co_2O$ with weak spin-orbit coupling, superconductivity remains highly sensitive to the external stimuli such as pressure.

## Introduction

The discovery and understanding of superconductivity (SC) with high superconducting transition temperatures ($T_c$) and large upper critical fields $B_{c2}(0)$ remain central challenges in the condensed matter physics. Materials with exceptionally high $B_{c2}(0)$ are technologically attractive for applications such as magnetic resonance imaging, nuclear magnetic resonance spectroscopy, and superconducting magnetic energy storage [1, 2]. Among them, a group of $\eta$-carbide-type superconductors including $Nb_4Rh_2C_{1-\delta}$ [3], $Ta_4Rh_2C_{1-\delta}$ [4], $Ti_4Ir_2O$ [5, 6], and $Zr_4Pd_2O$ [7] have attracted much attention because the $B_{c2}(0)$ values exceed the Pauli paramagnetic limit given as the $B_p^{BCS} \approx 1.84[T/K]T_c$ for a weak-coupling BCS superconductor, while the $T_c$ can be tuned by chemical composition, pressure [8, 9]. In addition, these superconductors exhibit excellent mechanical properties, including large bulk modulus and high hardness, making them promising candidates for magnet manufacturing [10].

The origin of large $B_{c2}(0)$ in $\eta$-carbide-type superconductors remains unclear and lacks a unified explanation. It is generally attributed to the interplay of several factors, including strong spin-orbit coupling (SOC), enhanced electron correlations, complex Fermi-surface topology, and possible disorder effects in polycrystalline that can each dominate in different $\eta$-carbide-type superconductors [8, 9, 11, 12]. Previous studies on the $Ti_4Ir_2O$ have revealed that pressure-driven Fermi-surface reconstruction near the $K$ point with strong SOC is responsible for large $B_{c2}(0)$ [9], which was also recently confirmed by the density functional theory (DFT) calculations and analytic modeling [12]. It indicated that the electronic states near the $X$ points exhibit strong SOC, leading to a vanishing effective $g$ factor and thus an enhanced Pauli limiting field in $Ti_4Ir_2O$ [12]. At the same time, its isostructural $Nb_4Rh_2C_{1-\delta}$ with weaker SOC also exhibits similar crossover of the $B_{c2}(0)$ associated with pressure-induced Fermi surface reconstruction [8]. This may support the universality of large $B_{c2}(0)$ in the $\eta$-carbide-type superconductors. However, it seems that the influence of electronic correlation and SOC on the larger $B_{c2}(0)$ has been ignored. How to fully reflect their contributions requires more candidate materials, especially single crystal materials. In this regard, the growth of single-crystal $\eta$-carbide-type superconductors and the elucidation of their intrinsic superconducting and normal-state properties are of particular importance to clarify the underlying physical mechanisms.

The $Ti_4X_2O$ series ($X$ = Co, Rh, Ir) crystallizes in the cubic $\eta$-carbide structure with a space group $Fd$-3$m$ (No. 227) [5, 6, 9, 10, 12]. As illustrated in Fig. 1(a), the Ti1(16c), Ti2 (48f), and $X$(32e) atoms form a $Ti_2Ni$-type framework, while the oxygen occupies the interstitial 16d Wyckoff sites. The superconducting properties of $Ti_4X_2O$ are readily tunable by chemical substitution or external pressure. For example, the density of states at the Fermi level, $N(E_F)$, decreases from 11.22 to 9.64 and 7.50 states eV$^{-1}$ per f.u. as the $X$ atom changes from Co to Rh to Ir. Correspondingly, the superconducting transition temperature $T_c$ increases from 2.7 K to 2.8 K and 5.3 K, while the upper critical field $B_{c2}(0)$ changes from 7.08 T to 5.15 T and 16.06 T, respectively [5, 6, 9, 10]. Notably, both $Ti_4Co_2O$ and $Ti_4Ir_2O$ exhibit $B_{c2}(0)$ values exceeding the Pauli limit $B_p^{BCS}$. Compared with the 4d-Rh and 5d-Ir compounds, the 3d-Co analogue features weaker SOC but stronger electron correlations. In this respect, the $Ti_4Co_2O$ serves as an ideal comparison system to disentangle the relative roles of SOC and electronic correlation in contributing larger $B_{c2}(0)$ in the $\eta$-carbide superconductors. Systematic investigations of the normal-state and superconducting properties of $Ti_4Co_2O$ under pressure are crucial for understanding the intrinsic SC of $\eta$-carbide-type superconductors.

In this work, we carried out systematically high-pressure (HP) experiments on the single-crystal $Ti_4Co_2O$ through electrical transport, X-ray diffraction (XRD) and DFT calculations. We find that both the $T_c$ and $B_{c2}(0)$ exhibit non-monotonic pressure dependence and two distinct superconducting regimes (e.g., a low-pressure phase, SC1 for $P < 12.3$ GPa and a high-pressure phase, SC2 phase for $P > 20.5$ GPa) were revealed by comparing their different normal-state/superconducting characteristics. The absence of structural transition in the synchrotron XRD suggests that the crossover between SC1 and SC2 is more likely associated with a pressure-induced electronic reconstruction. Its large bulk modulus of ~ 192 GPa and the enhanced $B_{c2}(0)$ values make $Ti_4Co_2O$ a promising candidate for a high-incompressibility alloy superconducting material.

**Experimental methods**

High-quality single-crystal $Ti_4Co_2O$ was synthesized by the arc melting method and annealed at high temperature. The reactants titanium powder (purity: 99.9%, Alfa Aesar), cobalt powder (purity: 99.99%, Strem Chemicals), and titanium dioxide powder (purity: 99.9%, Strem Chemicals) were weighed in stoichiometric ratios, mixed, and pressed into a pellet in a glovebox. The pellet was melted on a water-cooled copper plate in an arc furnace. The sample was molten 10 times, sealed in a quartz ampule, and then annealed in a furnace for 30 days at 1000 °C. Single crystals of $Ti_4Co_2O$ were obtained by mechanically crushing solidified spheres.

At ambient pressure (AP), electrical resistivity $\rho(T)$ was performed by a standard four-probe configuration on the Physical Property Measurement System (PPMS-9T, 1.8-300 K, Quantum Design). Magnetic susceptibility $\chi(T)$ was measured on the Magnetic Property Measurement System (MPMS-III, 1.8-300 K, Quantum Design). The temperature dependence of resistance $R(T)$ was carried out by using a standard four-probe method in a piston-cylinder pressure cell (PCC) up to 2.03 GPa and a commercial non-magnetic CuBe diamond anvil cell (DAC) of 300 $\mu$m anvil culet up to 69.7 GPa, respectively. For the PCC measurements, the liquid pressure transmitting medium (PTM), i.e., Daphene 7373 was used and the pressure were determined from the superconducting transition of Pb. For the DAC measurements, a rhenium gasket was pre-indented to a thickness of ~ 33 $\mu$m and then a 100 $\mu$m diameter hole was drilled by a laser drilling system. The metallic gasket was insulated with *c*-BN epoxy layer. A $Ti_4Co_2O$ flake with the size of ~ 97×27×12 $\mu m^3$ was loaded in the center of hole filled with soft KBr as the PTM. The real pressure was determined by the shift of the *R*1 fluorescence line of ruby below 30 GPa and the Raman spectrum of diamond up to 69.7 GPa, respectively. Low-temperature *R*(T) data was carried out in a $^4$He cryostat equipped with a 9 T superconducting magnet. Ultra-low temperature *R*(T) was conducted in magnetic field up to 18 T down to 300 mK in a top-loading $^3$He refrigerator. High pressure X-ray diffraction (HP-XRD) was collected at the BL15U1 station of Shanghai Synchrotron Radiation Facility (SSRF) with a wavelength of 0.6199 Å. The ruby was used for pressure calibration and the 4:1 methanol-ethanol as the PTM. Lattice parameters were obtained from the XRD patterns refinement by using the GSAS software [13].

The DFT calculations were carried out by using the QUANTUM ESPRESSO (QE) package [14-16]. We used the projector augmented-wave pseudopotentials with Perdew-Burke-Ernzerhof exchange-correlation density functional (PBEsol) [17], as implemented in the PSLIBRARY [18]. Before self-consistent calculations of electronic structures, the atomic positions were relaxed while setting the lattice parameters to the experimental ones. The charge density was calculated by using a Monkhorst-Pack grid of $7^3$ k points, while a denser grid of $21^3$ k points was used to calculate the density of states (DOS). The spin-orbit coupling (SOC) and the Co-3*d* electron correlation were not considered in the DFT calculations.

**Results and discussion**

In Figs.1(b)-(f), both $\rho(T)$ and $\chi(T)$ of $Ti_4Co_2O$ are plotted to revisit its normal-state and superconducting properties at AP. We can find, low-temperature $\rho(T)$ behaves a Fermi liquid behavior, e.g., $\rho(T) \approx 51.08\times10^{-6}\ \Omega$ cm + $1.59\times10^{-9}\ \Omega$ cm $K^{-1}T^2$. On cooling further, $\rho(T)$ starts to decrease at ~ 2.8 K and reaches zero at ~ 2.6 K, as shown in Fig. 1(b). For clarify, the $T_c^{onset}$ was determined as the temperature deviating from normal-state resistivity, and the $T_c^{zero}$ is defined as the zero-resistivity temperature, respectively. In Fig. 1(c), $\chi(T)$ was measured at a fixed filed of $B$ = 20 Oe in both zero-field-cooled (ZFC) and field-cooled (FC) modes, respectively. The estimated superconducting shielding volume $4\pi\chi(T)$ (2 K) is ≈ 1, indicating bulk superconducting state in $Ti_4Co_2O$; the superconducting transition temperature $T_c^M \approx$ 2.8 K determined by $\chi(T)$, is in good agreement with the value determined by $\rho(T)$. As shown in Fig. 1(c), $\chi(T)$ at $B$ = 1.0 T exhibits a typical Pauli paramagnetic behavior without long-range magnetic order. To get the $B_{c2}(0)$, we measured the $\rho(B)$ at fixed temperatures in Fig. 1(d) and the $\rho(T)$ under various fields in Fig. 1(e), respectively. Using the data, we then adopted the criteria of $T_c^{onset}$ and plotted the temperature dependence of $B_{c2}(T)$ in Fig. 1(f). The temperature dependence of the $B_{c2}(T)$ was well fitted by using the Werthamer-Helfand-Hohenberg (WHH) model in a clean limit [19-21], viz., $B_{c2}(T) = B_{c2}(0)\ [(1\text{-}t)–C_1(1\text{-}t)^2–C_2(1\text{-}t)^4]/0.73$ with $C_1$ = 0.135 and $C_2$ = 0.134, respectively. The extracted $B_{c2}(0)$ value is about 7.2 T, and larger than the $B_p^{BCS} \approx 1.84T_c \approx 5.09$ T.

To track the evolution of the $T_c$ under pressure, temperature dependence of resistance $R(T)$ was performed in a PCC up to 2.03 GPa and a DAC up to 69.7 GPa, respectively. The temperature range from 2 to 300 K is shown in Figs. 2(a)-(c). To better illustrate the pressure-induced evolution of $T_c$, we enlarge the low-temperature region, as shown in Figs. 2(d)-(f). In the PCC, the normal-state resistivity decreases with increasing pressure (Fig. 2(a)), while the $T_c$ increases monotonically from 2.7 K at 0.06 GPa to 2.9 K at 2.03 GPa (Fig. 2(d)). In the DAC, as shown in Figs. 2(b) and 2(e), the resistance continues to decrease, but the $T_c$ exhibits a nonmonotonic pressure dependence: the $T_c^{onset}$ initially increases from 2.8 K at AP to 3.4 K at 12.3 GPa, accompanying by a significant broadening of superconducting transition. With further increasing pressures, the $T_c^{onset}$ first reaches a minimum of approximately 3 K and then increases continuously above 20.5 GPa with a sharp superconducting transition, as seen in Figs. 2(c) and 2(f). Notably, the $T_c^{onset}$ continues to increase up to 69.7 GPa without showing any sign of saturation. These features indicate the emergence of new superconducting phase (e.g., SC2) above 20.5 GPa. For clarify, the pressure dependence of $T_c^{onset}$ and $T_c^{zero}$ were extracted and summarized in Fig. 3(a). Both $T_c^{onset}$ and $T_c^{zero}$ obtained from PCC and DAC almost coincide with each other. In detail, it rises monotonically at first with a pressure coefficient of $dT_c/dP \approx 0.034$ K/GPa, but then decreases around 10-20 GPa, and then increases with a $dT_c/dP \approx 0.023$ K/GPa, up to ~ 4.31 K at 69.7 GPa. These results reveal that the SC2 emerges above 20.5 GPa and its $T_c$ remains robust.

To gain insight into the evolution of the $T_c$, the normal-state low-$T$ resistance just above $T_c$ is analyzed by the power-law formula, i.e., $R = R_0 + AT^n$, where the $R_0$ is the residual resistance, the coefficient $A$ and the exponent $n$ are related to the density of state at Fermi level and the inelastic electron scattering, respectively. The polynomial fittings from $T_c$ to 30 K are summarized in Fig. 3(b) and Figs. 1S(a)-(b), which gives the evolution of $R_0$ and $n$ as a function of pressure. Specifically, the quadratic temperature coefficient $A$ was also extracted from a linear fit to the ($R$-$R_0$) vs $T^2$ plot in Figs. 1S(c)-(d). The resulting pressure evolution of these characteristic parameters is summarized in Fig. 3(c)-3(f). In Fig. 3(c), it is found that the $R/R_0$ decreases over the entire pressure range, but exhibits a small protrusion around 14.3-20.5 GPa, coinciding with the pressure interval where the $T_c$ drops abruptly. A similar anomaly is also reflected in the evolution of the coefficient $A$. As displayed in Fig. 3(d), the $A$ decreases monotonically for $P < 14.3$ GPa, then drops by nearly three orders of magnitude at 20.5 GPa, and remains nearly constant upon further compression. Within the Kadowaki-Woods framework, the $A$ is proportional to the square of the Sommerfeld coefficient and is therefore often taken to reflect the evolution of the density of states at the Fermi level, $N(E_F)$. In this sense, the overall reduction of the $A$ may indicate the suppression of $N(E_F)$ under pressure, although this trend appears opposite to the recovery of the $T_c$ in the high-pressure region of $Ti_4Co_2O$. More importantly, the exponent $n$ exhibits a pronounced pressure dependence. As shown in Fig. 3(f), the $n$ increases from 2 at AP to ≈ 4 above 20.5 GPa, and then remains nearly unchanged. This behavior differs from typical values of $n$ = 2(5) usually associated with the electron-electron (electron-phonon scattering). In $Ti_4Co_2O$, the $n \approx 4$ deviating from Fermi liquid under pressure indicate pressure-induced enhancement in the phonon scatterings and/or significant changes in the phonon spectra. Therefore, the evolution of $n$, together with the anomalies in the $R_0$ and $A$, are positively correlated with the changes of $T_c$, pointing to a clear modification of electronic structure and/or electron scattering mechanism under compression.

To further examine whether the pressure-induced changes in low-temperature transport are related to the phonons, we then estimated the Debye temperature $\Theta_D$ using Bloch-Grüneisen (B-G) analysis and the parallel resistance model, which is an important parameter for characterizing phonon spectra and lattice vibrations, as shown in Figs. S2(a)-S2(d). In the high-temperature B-G model, namely for $T > 0.5\Theta_D$ ($\Theta_D \approx 386$ K in $Ti_4Co_2O$), the resistance can be approximated as $R(T) \approx A'T/4M\Theta_D^2 \approx B$T in the B-G model [22, 23], where the $A$' and $M$ represents the characteristic parameter and the atomic mass, respectively. Based on this approximation, the pressure dependence of $\Theta_D(P)/\Theta_D^{AP}$ extracted from the high-temperature resistivity was summarized in Fig. 3(e). The ratio $\Theta_D(P)/\Theta_D^{AP}$ remains nearly constant at ∼ 1 for $P < 20.5$ GPa and then increases gradually to 1.23 at 69.7 GPa. By contrast, the $\Theta_D(P)/\Theta_D^{AP}$ values obtained from the parallel-resistor model [24, 25], which fits the full temperature range at each pressure, exhibit a markedly different pressure evolution. As shown in Fig. 3(e), this ratio increases monotonically from 1.0 at AP to 1.75 at 20.5 GPa, and then reaches saturation up to 69.7 GPa. The contrast between two analyses suggests that pressure has a stronger effect on the phonon-related transport response in low-temperature region than that inferred from high-temperature B-G approximation alone. In particular, the distinct change near 20.5 GPa is basically consistent with the anomalies in the $R_0$, $A$ and $n$, supporting that the crossover around this pressure involves not only changes in electronic structures but also substantial modification of phonon scattering.

Having identified pronounced anomalies in normal-state transport and phonon-related response near 20.5 GPa, we next examine whether the superconducting state exhibits similar pressure-induced crossover. As we know, the $B_{c2}(0)$ is one of fundamental parameters for superconductors, and its variation reflects superconducting pairing strengths and electron effective mass, and therefore provides an additional perspective on the pressure evolution of the $T_c$ in $Ti_4Co_2O$. The temperature-dependent resistance under different magnetic fields is shown in Fig. 3S. Here, in Figs. 4(a)-(b), we take the $T_c$ using the criteria of $T_c^{onset}$ and fit the $B_{c2}(0)$ by using the WHH model as above. The resulting pressure dependence of $B_{c2}(0)$, together with the Pauli limit $B_p^{BCS}$ for comparison, is summarized in Fig. 4(c). As can be seen, the $B_{c2}(0)$ exhibits a clear nonmonotonic pressure dependence: it first increases from 7.2 T at AP to 12.3 T at 4.2 GPa, then decreases to 7.35 T at 33 GPa, and finally increase again to 8.74 T at 69.7 GPa, quiet similar to the pressure evolution of $T_c$. Notably, the $B_{c2}(0)$ remains larger than $B_p^{BCS}$ within our pressure range. This behavior differs from those of isostructural superconductors $Ti_4Ir_2O$ [9] and $Nb_4Rh_2C_{1-\delta}$ [8], where the $B_{c2}(0)$ shows a smoother crossover from well beyond to less than the $B_p^{BCS}$, which was attributed to Fermi surface reconstruction. For comparison, we extracted the normalized slope $(1/T_c)dB_{c2}/dT|T = T_c$. Within a single-band picture, it can serve as an indicator of the electron effective mass. As shown in Fig. 4(c), its pressure dependence is similar to that of $B_{c2}(0)$, reaching a maximum near 5 GPa and then decreasing monotonically with further increasing pressure. These results indicate that the electron effective mass has undergone significant changes, consistent with the crossover inferred from the transport analysis.

To further investigate possible structural transition, HP-XRD was performed up to 55.8 GPa. As shown in Fig. 5(a), no splitting of diffraction peaks and new peak formation were observed, indicating the absence of structural transition below 55.8 GPa. The Rietveld refinements using GSAS, shown in Figs. 5(b)-(c), further reveal that both lattice parameter $a$ and volume $V$ decrease smoothly with increasing pressure. Their pressure dependences, together with the corresponding normalized quantities $a/a_0$ and $V/V_0$, are summarized in Fig. 5(d). At 55.8 GPa, the $a$ and $V$ are reduced by about 6.2% and 17%, respectively, relative to their values at AP, but with no detectable anomaly around 20 GPa. The high-pressure fitting of $V(P)$ to the Birch-Murnaghan (B-M) equation ($B' = 4$) yields the $B_0 \approx 192$ GPa, $V_0 \approx 1437.16$ Å$^3$. This bulk modulus is substantially larger than that of the common alloy superconductors (eg., $B_0 \approx 100$ GPa) and close to other $\eta$-carbide-type superconductors (eg., $B_0 \approx 251.8(1)$ GPa for $Ti_4Ir_2O$ [9]; $B_0 \approx 278.6$ GPa for $Nb_4Rh_2C_{1-\delta}$ [8], etc.), as well as the typical perovskite oxides ($B_0 \approx 240$-$250$ GPa [26-28]). Therefore, the smooth structural evolution with pressure rules out structural transition for the anomalous behavior of $T_c$ and $B_{c2}(0)$, but instead supports a transition of electronic states.

To further verify such hypothesis, we have performed the DFT calculations on the electronic structures of $Ti_4Co_2O$ and the results were summarized in Fig. 6 and Figs.4S-5S. We found, for each pressure, the Ti and Co atoms mainly contribute to the DOS near the $E_F$. With the pressures increasing up to 55.8 GPa, five bands crossing $E_F$ remain unchanged, creating five distinct sheets of Fermi surfaces (Fig. 5S). The topology of Fermi surfaces doesn't change except that their curvatures become prominent upon increasing pressures. Indeed, band structures (Fig. 4S) reveal that the energy dispersion of these five bands becomes more significant, leading to an overall decrease of electron effective mass $m^*$, basically consistent with the above results. Although there was no significant change in the contributions of the five energy bands to DOS at the $E_F$, the contribution of different bands to the electron-phonon coupling strength may be different, resulting in the multi band effect responsible for the non monotonic evolution of $T_c$. Fig. 6(a)-(b) shows the band structure and the DOS of $Ti_4Co_2O$ at AP. We found, the DOS at ~ 0.25 ± 0.02 eV is very low, indicating a pseudogap feature; the $E_F$ is just located at a local minimum of DOS. In Fig. 6(c), with increasing pressure, the DOS peak right below the $E_F$ gradually shift to higher energies, causing the nonmonotonic pressure dependence of the $N(E_F)$. In detail, as shown in Fig. 6(d), the $N(E_F)$ first monotonously decreases, reaching a minimum at 20 GPa, and then gradually increases. This trend is consistent with the decrease in the $T_c$ at low pressures, showing a positive correlation between them. This nonmonotonic trend arises from the fact that the $E_F$ of $Ti_4Co_2O$ is located at a local minimum of DOS, different from the isostructural $Ti_4Ir_2O$ [9] and $Nb_4Rh_2C_{1-\delta}$ [8]. Besides, the increase of $T_c$ can be attributed to lattice stiffening under pressure considering that the electron-phonon coupling is proportional to the logarithmic average phonon frequency $\omega_{\ln}$ [29]. In this regard, more studies involving the electron-phonon coupling and structural instability are required to understand the evolution of $T_c$ in $Ti_4Co_2O$ under pressure.

Finally, it is instructive to compare $Ti_4Co_2O$ with the related $\eta$-carbide superconductors $Ti_4Ir_2O$ and $Nb_4Rh_2C_{1-\delta}$. In $Ti_4Ir_2O$, pressure-driven Fermi-surface reconstruction together with strong SOC has been proposed to enhance the Pauli limiting field (via a strongly reduced effective $g$ factor near the $X$ points), leading to a characteristic pressure evolution of $B_{c2}(0)$ [9, 12]. In $Nb_4Rh_2C_{1-\delta}$, a similar pressure-induced crossover of $B_{c2}(0)$ relative to the weak-coupling Pauli limit has also been discussed in terms of an electronic reconstruction [8]. By contrast, $Ti_4Co_2O$ displays two superconducting regimes, namely, low-pressure SC1 phase (< 12.3 GPa) and high-pressure SC2 phase (> 20.5 GPa), which are separated by an intermediate crossover region where the superconducting transition broadens largely. This crossover is accompanied by systematic changes in electrical transport, including the exponent $n$ in $R = R_0+AT^n$ from ~ 2 at low pressures to ~ 4 at high pressures, as well as the anomalies in $R_0$ and $A$. At the same time, the $B_{c2}(0)$ value of $Ti_4Co_2O$ remains above $B_p^{BCS}$ throughout the entire pressures, in contrast to the smooth crossover of $B_{c2}(0)$ from beyond to below the Pauli limit in isostructural $Ti_4Ir_2O$ and $Nb_4Rh_2C_{1-\delta}$.

We may also note that the superconducting transition broadening in ~ 10-20 GPa maybe amplified by pressure inhomogeneity and non-hydrostatic stress in the DAC; nevertheless, the recovery of a sharp transition above ~20.5 GPa together with concurrent and reproducible changes in the normal-state fitting parameters suggests that the SC1 to SC2 crossover cannot be simply attributed to extrinsic effects alone. Moreover, high-pressure XRD reveals no structural transition up to 55.8 GPa. Instead, our DFT calculations show that the DOS peak gradually moves from below to above $E_F$, yielding a nonmonotonic behavior of $N(E_F)$ with a minimum near ~ 20 GPa that correlates with the evolution of $T_c$. Therefore, the appearance of two superconducting regimes in $Ti_4Co_2O$ likely reflects modification of DOS and/or pronounced change in electron-phonon scattering under pressures. More experimental investigations are required to uncover more physical mechanisms of $\eta$-carbide superconductors.

**Conclusion**

We reported the evolution of the $T_c$ and $B_{c2}(0)$ in $Ti_4Co_2O$ under pressures up to 69.7 GPa. Our results reveal that both the $T_c$ and $B_{c2}(0)$ exhibits non-monotonic pressure dependence, thus identifying two different superconducting regimes. This feature is confirmed by comparing normal states/superconducting characteristics. The $B_{c2}(0)$ shows a dome-like pressure dependence with its maximum at 5 GPa, and remains above the Pauli-limit $B_p^{BCS}$ up to 69.7 GPa. Combined with the synchrotron X-ray diffraction measurements and the DFT calculations, the nonmonotonic evolution of $T_c$ may be related to the change of electronic structures. Comparisons with $Ti_4Ir_2O$ and $Nb_4Rh_2C_{1-\delta}$ further indicate that the location of Fermi level may play a central role in shaping the superconducting behavior of $Ti_4Co_2O$. Overall, our findings establish the $Ti_4Co_2O$ as a candidate $\eta$-carbide superconductor in which electronic reconstruction is responsible for the emergence of different superconducting regimes under pressures.

**Acknowledgements**

This work is supported by the National Key Research and Development Program of China (2023YFA1406100, 2023YFA1607402, 2021YFA1400200, 2021YFA1401800), the National Natural Science Foundation of China (12025408, 11921004, 11834016, 12074414), the Strategic Priority Research Program of CAS (XDB25000000, XDB33000000), K.C. Wong Education Foundation (GJTD-2020-01), the Users with Excellence Program of Hefei Science Center CAS (2021HSC-UE008), the Youth Promotion Association of CAS (2018010) and the Outstanding member of Youth Promotion Association of CAS (Y2022004). HPXRD measurements were performed at the BL15U1 station of the Shanghai Synchrotron Radiation Facility (SSRF). This work is partially supported by the CAC station of Synergic Extreme Condition User Facility (SECUF).

**Figure Captions:**

**FIG. 1.** (a) Crystal structure of $Ti_4Co_2O$. (b) The electrical resistivity $\rho(T)$ at ambient pressure. The red line represents low-temperature Fermi liquid behavior, $\rho(T) \approx 51.08\times10^{-6}\ \Omega$ cm + $1.59\times10^{-9}\ \Omega$ cm $K^{-1}T^2$. The inset shows the enlarged view of superconducting transition, with the $T_c^{onset}$ and $T_c^{zero}$ as the onset of transition and the zero-resistivity temperature, respectively. (c) The magnetic susceptibility $\chi(T)$ at $B$ = 1 T. The inset shows the superconducting shielding volume $4\pi\chi(T)$ at $B$ = 20 Oe in both zero-field-cooled (ZFC) and field-cooled (FC) modes, respectively. The $T_c^M$ is the superconducting transition temperature determined by $\chi(T)$. (d) $\rho(H)$ under various temperatures (0.35-1.66 K). (e) $\rho(T)$ under various magnetic fields. (f) The $B_{c2}(T)$ and the mathematical fitting results by the Werthamer-Helfand-Hohenberg (WHH) model in a clean limit. The dashed red line represents the Pauli limit $B_P^{BCS} \approx 1.84T_c \approx 5.09$ T.

**FIG. 2.** $R(T)$ data under various pressures: (a) 0.06-2.03 GPa in a PCC, 1.5-300 K; (b) 1.5-20.5 GPa in a DAC, 1.5-300 K; (c) 20.5-69.7 GPa in a DAC, 1.5-300 K; (d) 0.06-2.03 GPa in a PCC, 2.2-3 K; (e) 1.5-20.5 GPa in a DAC, 2-3.5 K; (f) 20.5-69.7 GPa in a DAC, 2-4.5 K; The arrows in (d)-(f) show the $T_c^{onset}$.

**FIG. 3.** (a) Pressure dependence of the $T_c^{onset}$ and $T_c^{zero}$ in the PCC and DAC; (b) The normal-state low-$T$ resistance just above $T_c$ is analyzed by the power-law formula, i.e., $R = R_0+AT^n$, where the $R_0$ is the residual resistance, the coefficient $A$ and the exponent $n$ are related to the density of state at Fermi level and the inelastic electron scattering, respectively. Pressure dependence of characteristic parameters: (c) the $R_0/R_0$(AP), (d) the $A$ value and its first derivative $dA/dT$, (e) the $\Theta_D/\Theta_D$(AP) by using the Bloch-Grüneisen (B-G) formula and the parallel-resistor model, respectively, and (f) the exponent $n$, respectively.

**FIG. 4.** The $B_{c2}(T)$-$T$ data under various pressures: (a) from ambient pressure to 20.5 GPa, (b) 24.7- 69.7 GPa. The solid lines represent the mathematical fittings by using the WHH model. Pressure dependence of several characteristic parameters: (c) the $B_{c2}(0)$ and $B_P^{BCS}$, (d) the $(-dB_{c2}/dT)|_{T=T_c}$.

**FIG. 5.** (a) The XRD patterns of $Ti_4Co_2O$ under various pressures. Pressure dependence of characteristic parameters: (b) lattice parameter $a$, (c) the volume $V$, (d) the relative change $a/a_0$ and $V/V_0$. The solid lines in (c) shows the Birch-Murnaghan (B-M) fitting ($B'$=4 (fixed)).

**FIG. 6.** (a) The calculated electronic band structure of $Ti_4Co_2O$ at ambient pressure. (b) The calculated density of states (DOS) with projections on different atoms. (c) The evolution of DOS near the Fermi level ($E_F$) upon pressures. (d) Pressure dependence of the DOS at the $E_F$. The dash line indicates the changing trend.

**Figure 1**

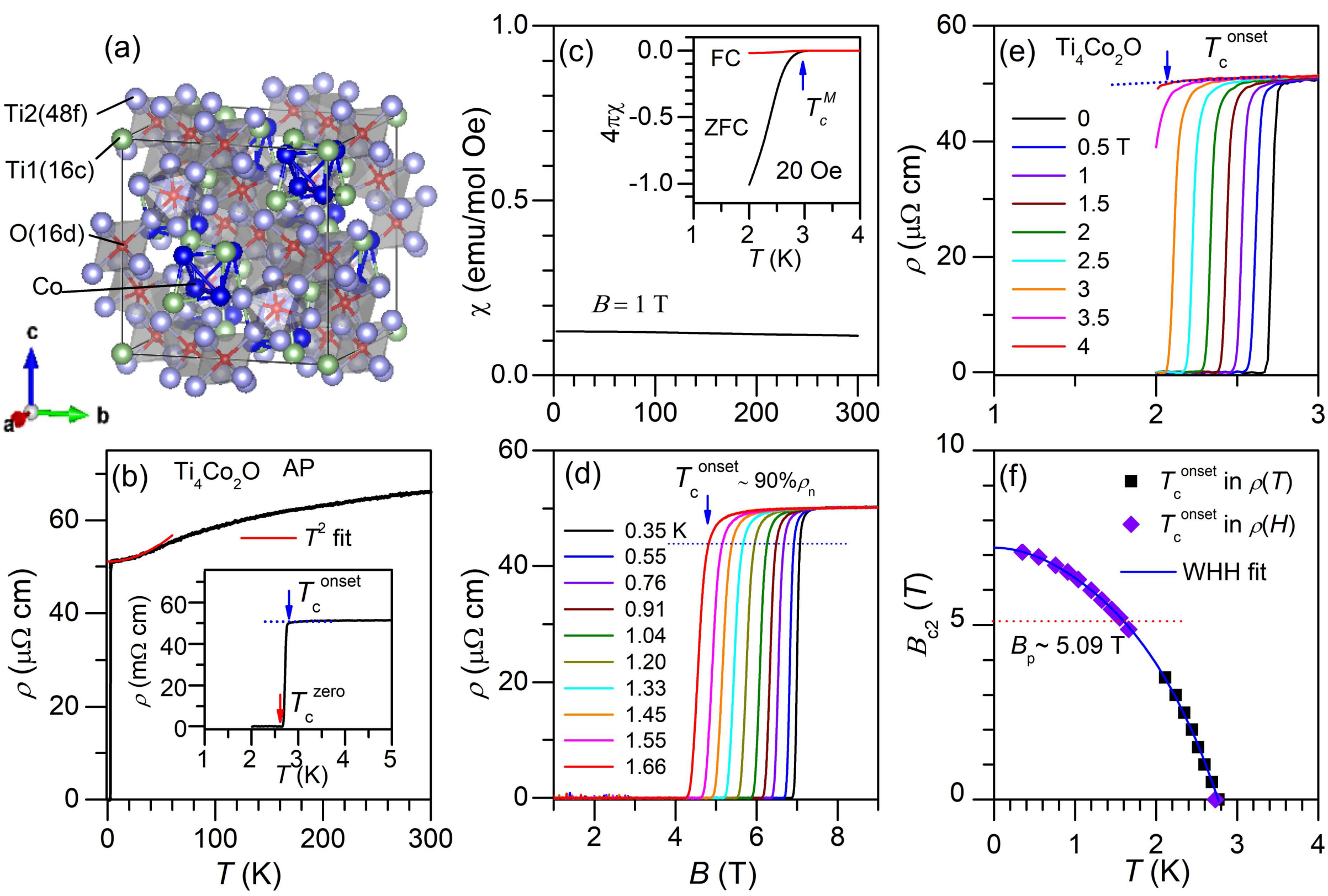

(a)
Ti2(48f)
Ti1(16c)
O(16d)
Co
c
a
b
(b)
Ti4Co2O AP
T2 fit
ρ (μΩ cm)
T (K)
Tc onset
Tc zero
ρ (mΩ cm)
T (K)
(c)
FC
ZFC
Tc M
20 Oe
4πχ
T (K)
B = 1 T
χ (emu/mol Oe)
(d)
Tc onset ~ 90%ρn
0.35 K
0.55
0.76
0.91
1.04
1.20
1.33
1.45
1.55
1.66
ρ (μΩ cm)
B (T)
(e)
Ti4Co2O
Tc onset
0
0.5 T
1
1.5
2
2.5
3
3.5
4
ρ (μΩ cm)
(f)
Tc onset in ρ(T)
Tc onset in ρ(H)
WHH fit
Bc2 (T)
Bp ~ 5.09 T
T (K)

**Figure 2**

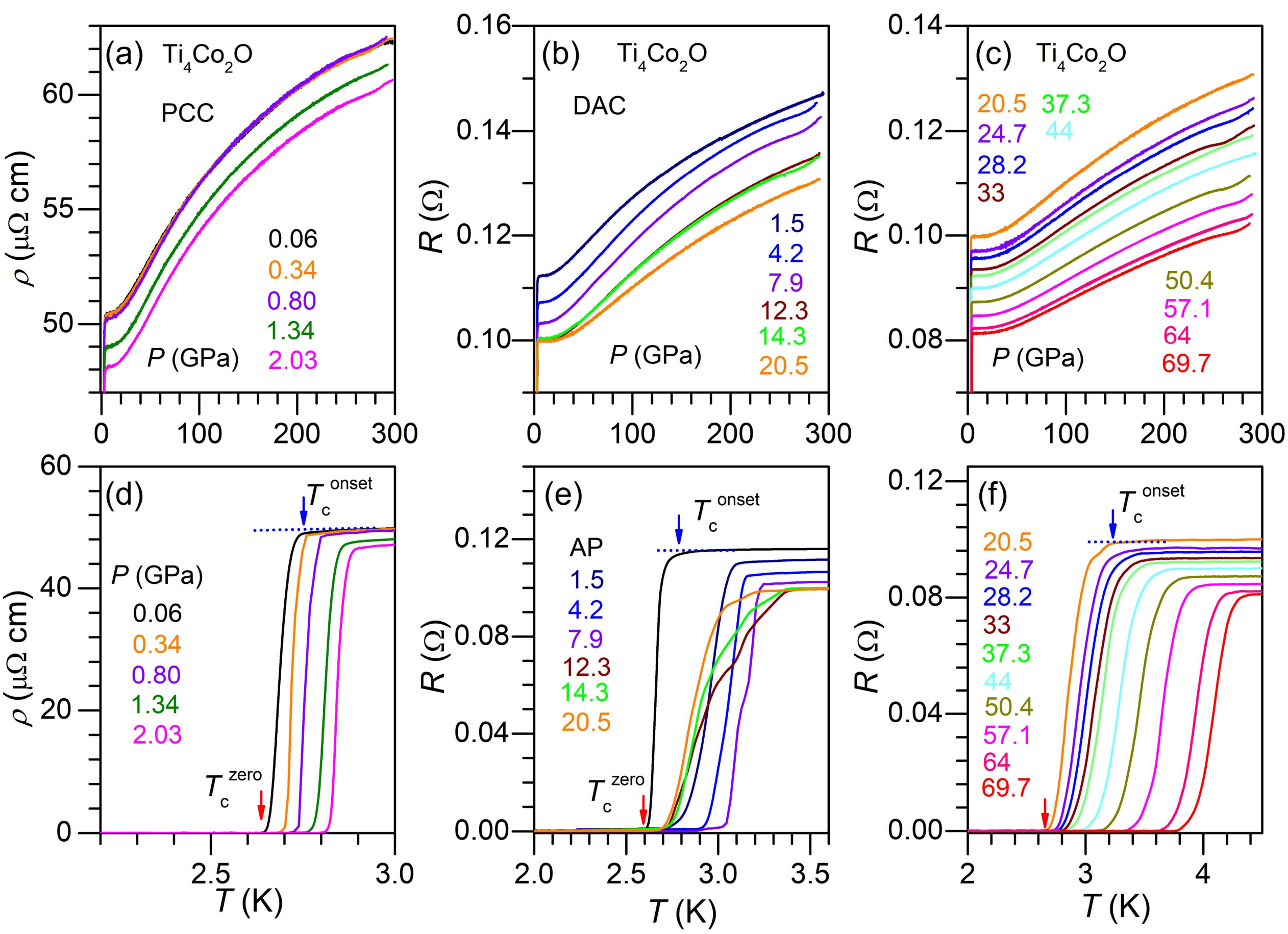

(a) Ti4Co2O
PCC
ρ (μΩ cm)
0.06
0.34
0.80
1.34
2.03
P (GPa)
(b) Ti4Co2O
DAC
R (Ω)
1.5
4.2
7.9
12.3
14.3
20.5
P (GPa)
(c) Ti4Co2O
20.5 37.3
24.7 44
28.2
33
50.4
57.1
64
69.7
P (GPa)
R (Ω)
(d)
Tc onset
P (GPa)
0.06
0.34
0.80
1.34
2.03
Tc zero
ρ (μΩ cm)
T (K)
(e)
Tc onset
AP
1.5
4.2
7.9
12.3
14.3
20.5
Tc zero
R (Ω)
T (K)
(f)
Tc onset
20.5
24.7
28.2
33
37.3
44
50.4
57.1
64
69.7
R (Ω)
T (K)

**Figure 3**

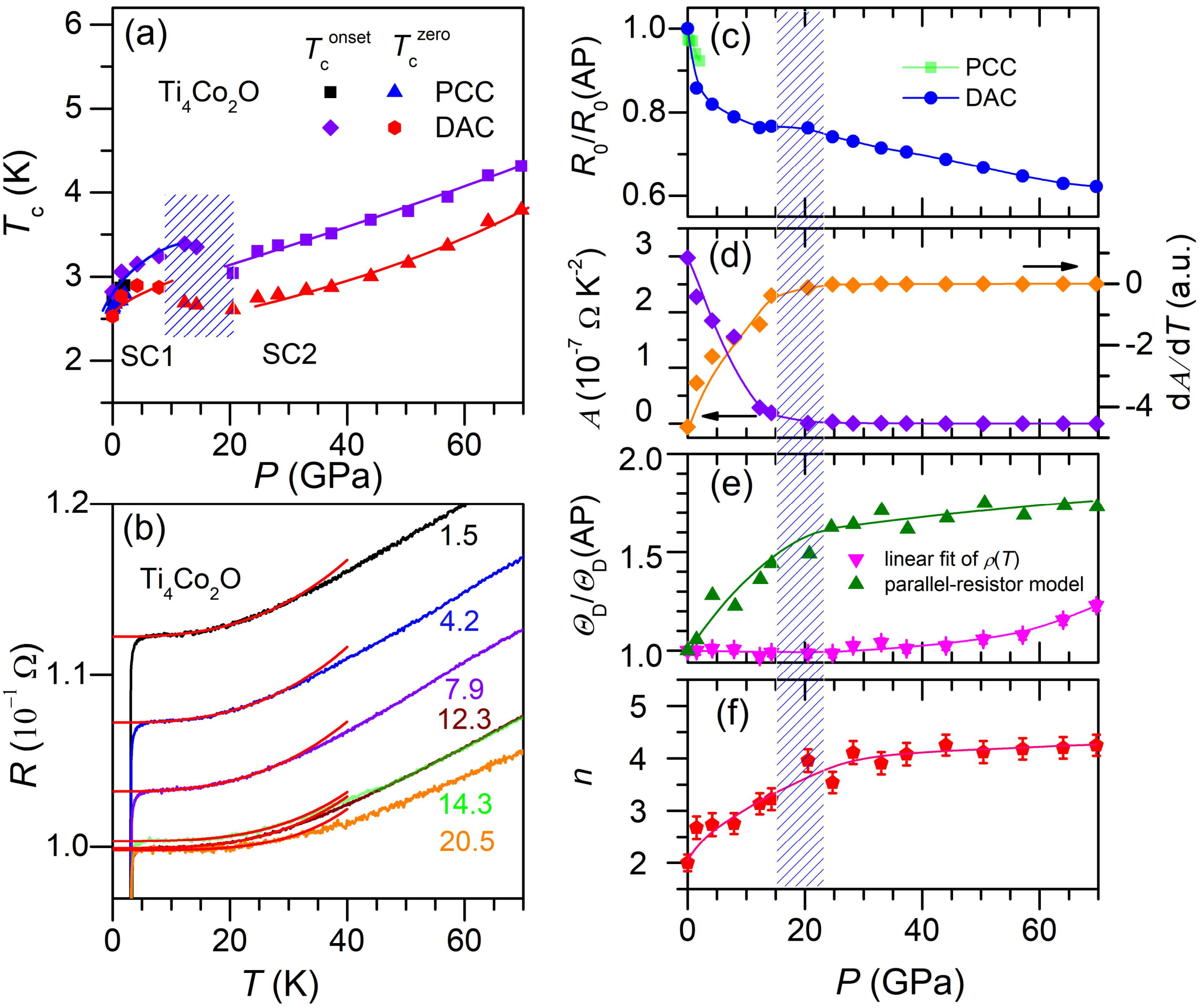

(a)
Ti4Co2O
Tc onset
Tc zero
PCC
DAC
SC1
SC2
Tc (K)
P (GPa)
(b)
Ti4Co2O
1.5
4.2
7.9
12.3
14.3
20.5
R (10^-1 Ω)
T (K)
(c)
PCC
DAC
R0/R0(AP)
(d)
A (10^-7 Ω K^-2)
dA/dT (a.u.)
(e)
linear fit of ρ(T)
parallel-resistor model
ΘD/ΘD(AP)
(f)
n
P (GPa)

**Figure 4**

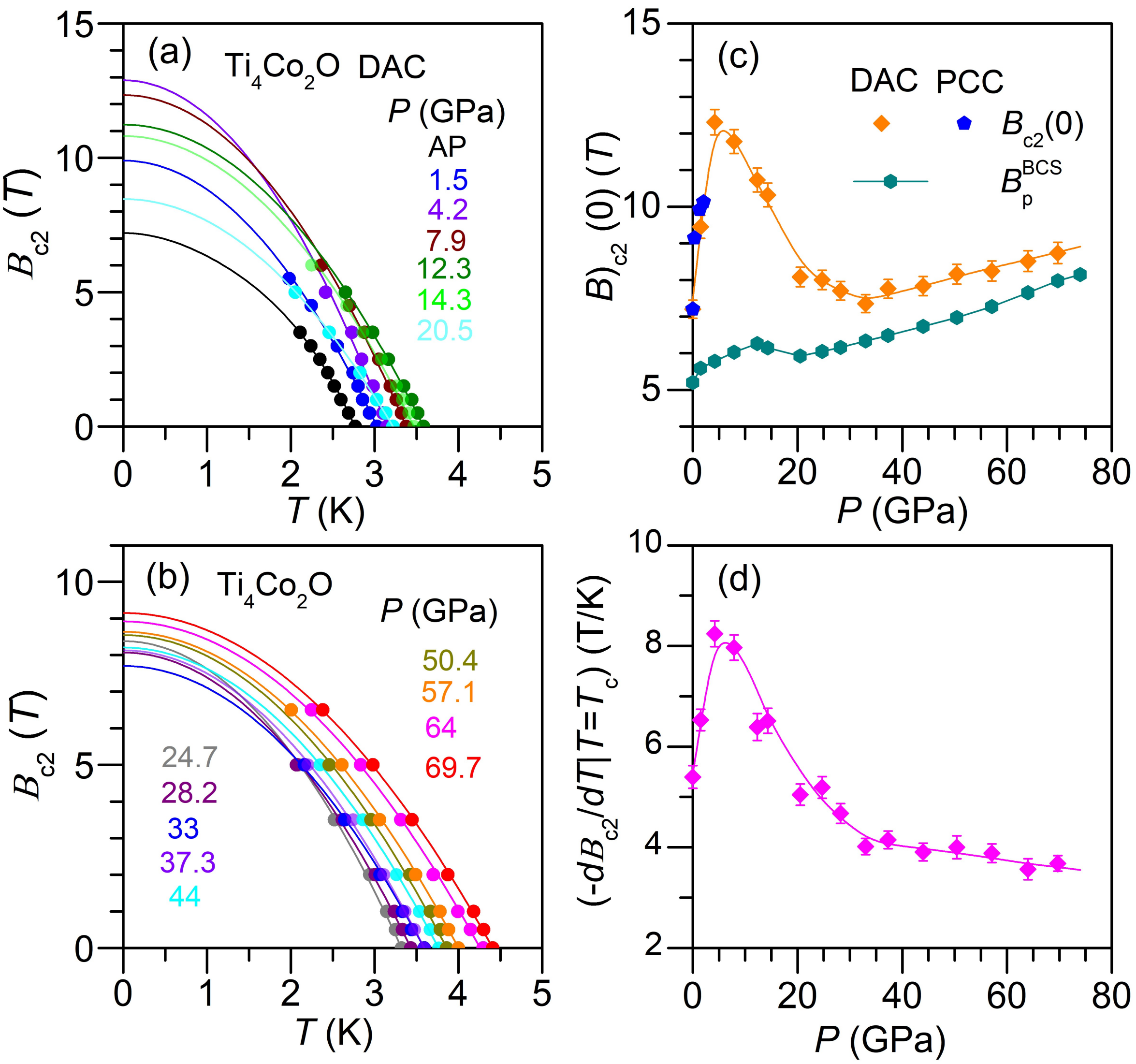

(a)
Ti4Co2O DAC
P (GPa)
AP
1.5
4.2
7.9
12.3
14.3
20.5
Bc2 (T)
T (K)
(b)
Ti4Co2O
P (GPa)
50.4
57.1
64
69.7
24.7
28.2
33
37.3
44
Bc2 (T)
T (K)
(c)
DAC PCC
Bc2(0)
Bp BCS
Bc2(0) (T)
P (GPa)
(d)
(-dBc2/dT|T=Tc) (T/K)
P (GPa)

**Figure 5**

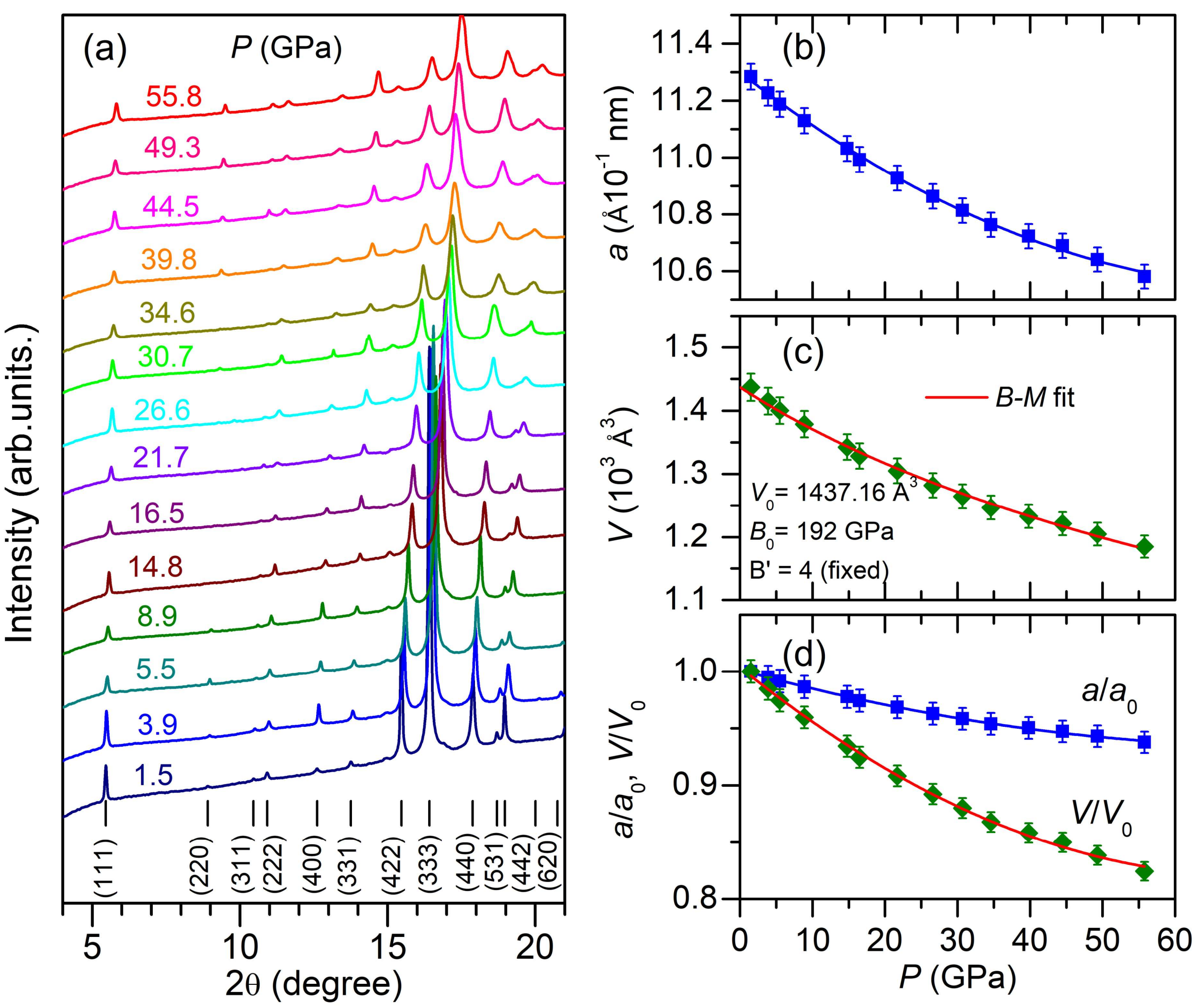
(a)
P (GPa)
55.8
49.3
44.5
39.8
34.6
30.7
26.6
21.7
16.5
14.8
8.9
5.5
3.9
1.5
(111)
(220)
(311)
(222)
(400)
(331)
(422)
(333)
(440)
(531)
(442)
(620)
Intensity (arb.units.)
2θ (degree)
(b)
a (Å10^-1 nm)
(c)
B-M fit
V (10^3 Å^3)
V0= 1437.16 Å^3
B0= 192 GPa
B' = 4 (fixed)
(d)
a/a0
V/V0
a/a0, V/V0
P (GPa)

**Figure 6**

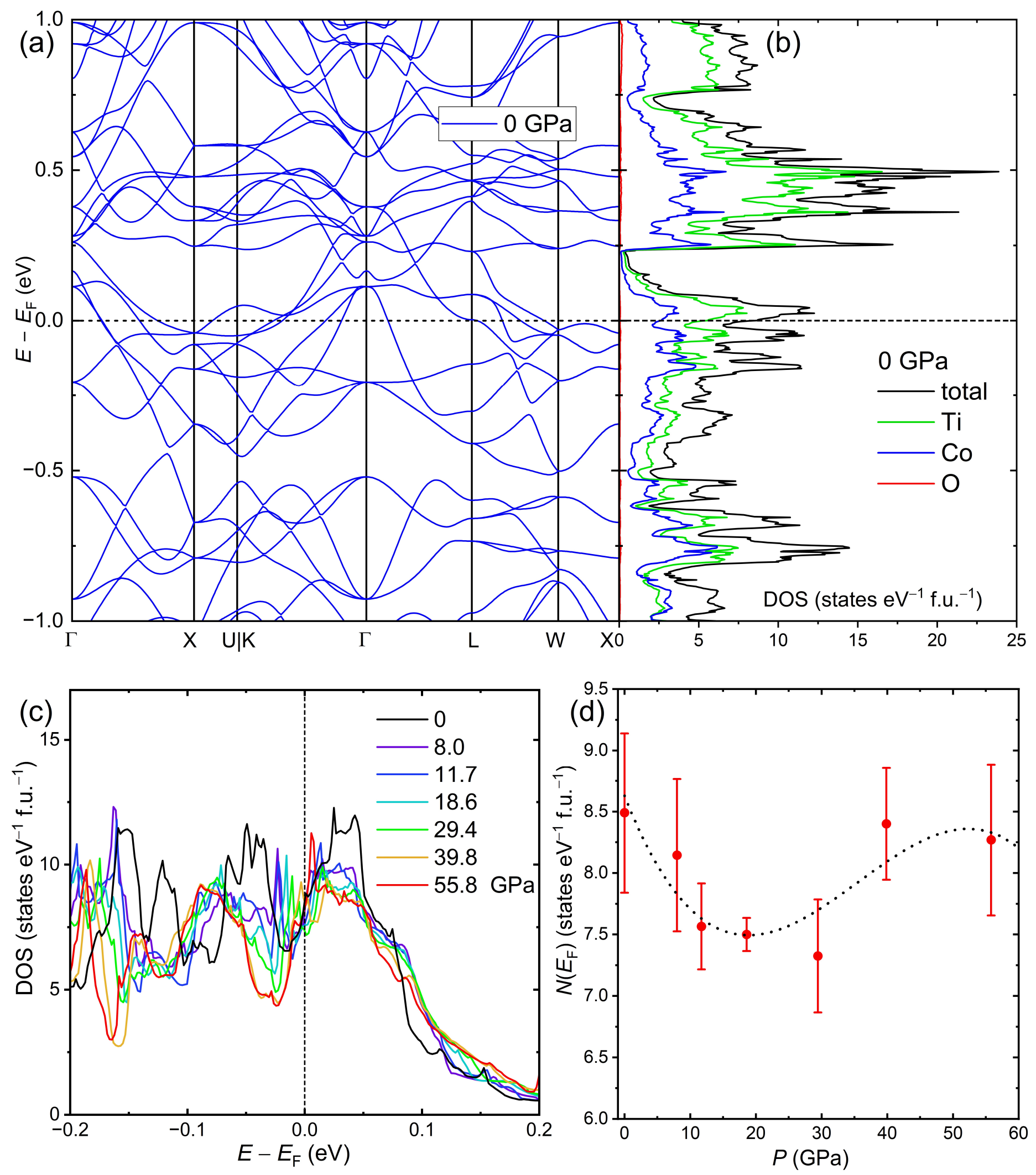

(a)
(b)
0 GPa
E − E_F (eV)
Γ X U|K Γ L W X
0 GPa
total
Ti
Co
O
DOS (states eV⁻¹ f.u.⁻¹)
(c)
0
8.0
11.7
18.6
29.4
39.8
55.8 GPa
DOS (states eV⁻¹ f.u.⁻¹)
E − E_F (eV)
(d)
N(E_F) (states eV⁻¹ f.u.⁻¹)
P (GPa)

## References

[1] H. Hosono, A. Yamamoto, H. Hiramatsu, and Y. Ma, Recent advances in iron-based superconductors toward applications, Mater. Today **21**, 278 (2018).

[2] C. Yao and Y. Ma, Superconducting materials: Challenges and opportunities for large-scale applications, iScience **24**, 102541 (2021).

[3] K. Ma, K. Gornicka, R. Lefevre, Y. Yang, H. M. Ronnow, H. O. Jeschke, T. Klimczuk, and F. O. von Rohr, Superconductivity with High Upper Critical Field in the Cubic Centrosymmetric $\eta$-Carbide $Nb_4Rh_2C_{1-\delta}$, ACS Mater Au **1**, 55 (2021).

[4] K. Ma, S. López-Paz, K. Gornicka, H. O. Jeschke, T. Klimczuk, and F. O. von Rohr, Discovery of the Type-II Superconductor $Ta_4Rh_2C_{1-\delta}$ with a High Upper Critical Field, Phys. Rev. Res. **7**, 023147 (2025).

[5] K. Ma, R. Lefèvre, K. Gornicka, H. O. Jeschke, X. Zhang, Z. Guguchia, T. Klimczuk, and F. O. von Rohr, Group-9 Transition-Metal Suboxides Adopting the Filled-$Ti_2Ni$ Structure: A Class of Superconductors Exhibiting Exceptionally High Upper Critical Fields, Chem. Mater. **33**, 8722 (2021).

[6] B.-B. Ruan, M.-H. Zhou, Q.-S. Yang, Y.-D. Gu, M.-W. Ma, G.-F. Chen, and Z.-A. Ren, Superconductivity with a Violation of Pauli Limit and Evidences for Multigap in $\eta$-Carbide Type $Ti_4Ir_2O$, Chin. Phys. Lett. **39**, 027401 (2022).

[7] Y. Watanabe, A. Miura, C. Moriyoshi, A. Yamashita, and Y. Mizuguchi, Observation of superconductivity and enhanced upper critical field of η-carbide-type oxide $Zr_4Pd_2O$, Sci. Rep. **13**, 22458 (2023).

[8] L. Shi, K. Ma, B. Ruan, N. Wang, J. Hou, P. Shan, P. Yang, J. Sun, G. Chen, Z. Ren, F. O. von Rohr, B. Wang, and J. Cheng, Nonmonotonic superconducting transition temperature and large bulk modulus in the alloy superconductor $Nb_4Rh_2C$, Phys. Rev. B **110**, 214520 (2024).

[9] L. Shi, B. Ruan, P. Yang, N. Wang, P. Shan, Z. Liu, J. Sun, Y. Uwatoko, G. Chen, Z. Ren, B. Wang, and J. Cheng, Pressure-driven evolution of upper critical field and Fermi surface reconstruction in the strong-coupling superconductor $Ti_4Ir_2O$, Phys. Rev. B **107**, 174525 (2023).

[10] L. Shi, K. Ma, J. Hou, P. Ying, N. Wang, X. Xiang, P. Yang, X. Yu, H. Gou, and J. Sun, Synergetic Enhancement of Hardness and Toughness in New Superconductors $Ti_2Co$ and $Ti_4Co_2O$, Chin. Phys. Lett. **42**, 067302 (2025).

[11] D. Das, K. Ma, J. Jaroszynski, V. Sazgari, T. Klimczuk, F. O. von Rohr, and Z. Guguchia, $Ti_4Ir_2O$: A time reversal invariant fully gapped unconventional superconductor, Phys. Rev. B **110**, 174507 (2024).

[12] H. Wu, T. Shishidou, M. Weinert, and D. F. Agterberg, Large critical fields in superconducting $Ti_4Ir_2O$ from spin-orbit coupling, Phys. Rev. B **111**, 184506 (2025).

[13] B. H. Toby, A Graphical User Interface for GSAS, J. Appl. Crystallogr. **34**, 210 (2001).

[14] P. Giannozzi, O. Andreussi, T. Brumme, O. Bunau, M. Buongiorno Nardelli, M. Calandra, R. Car, C. Cavazzoni, D. Ceresoli, M. Cococcioni, et al, Advanced

capabilities for materials modelling with Quantum ESPRESSO, J. Phys.: Condens. Matter. **29**, 465901 (2017).

[15] P. Giannozzi, S. Baroni, N. Bonini, M. Calandra, R. Car, C. Cavazzoni, D. Ceresoli, G. L. Chiarotti, M. Cococcioni, I. Dabo, et al, QUANTUM ESPRESSO: a modular and open-source software project for quantum simulations of materials, J. Phys.: Condens.Matter. **21**, 395502 (2009).

[16] P. Giannozzi, O. Baseggio, P. Bonfà, D. Brunato, R. Car, I. Carnimeo, C. Cavazzoni, S. De Gironcoli, P. Delugas, and F. Ferrari Ruffino, Quantum ESPRESSO toward the exascale, J. Chem. Phys. **152**, 154105 (2020).

[17] J. P. Perdew, A. Ruzsinszky, G. I. Csonka, O. A. Vydrov, G. E. Scuseria, L. A. Constantin, X. Zhou, and K. Burke, Restoring the Density-Gradient Expansion for Exchange in Solids and Surfaces, Phys. Rev. Lett. **100**, 136406 (2008).

[18] A. Dal Corso, Pseudopotentials periodic table: From H to Pu, Comput. Mater. **95**, 337 (2014).

[19] E. Helfand and N. R. Werthamer, Temperature and Purity Dependence of the Superconducting Critical Field, $H_{c2}$. II, Phys. Rev. **147**, 288 (1966).

[20] T. P. Orlando, E. J. McNiff, S. Foner, and M. R. Beasley, Critical fields, Pauli paramagnetic limiting, and material parameters of $Nb_3Sn$ and $V_3Si$, Phys. Rev. B **19**, 4545 (1979).

[21] N. R. Werthamer, E. Helfand, and P. C. Hohenberg, Temperature and Purity Dependence of the Superconducting Critical Field, $H_{c2}$. III. Electron Spin and Spin-Orbit Effects, Phys. Rev. **147**, 295 (1966).

[22] D. Cvijović, The Bloch-Gruneisen function of arbitrary order and its series representations, Theor. Math. Phys. **166**, 37 (2011).

[23] F. Bloch, Zum elektrischen Widerstandsgesetz bei tiefen Temperaturen, Zeitschrift für Physik **59**, 208 (1930).

[24] Arushi, K. Motla, A. Kataria, S. Sharma, J. Beare, M. Pula, M. Nugent, G. M. Luke, and R. P. Singh, Type-I superconductivity in single-crystal $Pb_2Pd$, Phys. Rev. B **103**, 184506 (2021).

[25] M. Li, D. Zhang, J. Han, Y. Fang, N. Li, X. Liu, B. Wang, L. Yan, F. Chen, Y.-L. Li, and W. Yang, Pressure-tuning structural and electronic transitions in semimetal CoSb, Phys. Rev. B **104**, 054511 (2021).

[26] Y. Ma and R. Aksoy, Dependence of band structures on stacking and field in layered graphene, Solid State Commun. **142**, 376 (2007).

[27] D. Valim, A. G. Souza Filho, P. T. C. Freire, S. B. Fagan, A. P. Ayala, J. Mendes Filho, A. F. L. Almeida, P. B. A. Fechine, A. S. B. Sombra, J. Staun Olsen, and L. Gerward, Raman scattering and x-ray diffraction studies of polycrystalline $CaCu_3Ti_4O_{12}$ under high-pressure, Phys. Rev. B **70**, 132103 (2004).

[28] A. S. Verma and A. Kumar, Bulk modulus of cubic perovskites, J. Alloys Compd. **541**, 210 (2012).

[29] P. B. Allen and R. C. Dynes, Transition temperature of strong-coupled superconductors reanalyzed, Phys. Rev. B **12**, 905 (1975).